\documentstyle[aps]{revtex} 
\begin{document}
\twocolumn
\title{The null energy condition in dynamic wormholes}
\author{
David Hochberg$^{+}$ and Matt Visser$^{++}$
}
\address{
$^{+}$Laboratorio de Astrof\'{\i}sica Espacial y F\'{\i}sica
Fundamental, Apartado 50727, 28080 Madrid, Spain\\
$^{++}$Physics Department, Washington University,
Saint Louis, Missouri 63130-4899, USA\\
}
\date{17 February 1998; Revised 30 April 1998}
\maketitle

\def\Tr{\hbox{\rm Tr}}

{\small {\em Abstract:} We extend previous proofs that violations
of the null energy condition (NEC) are a generic and universal
feature of traversable wormholes to completely non-symmetric
time-dependent wormholes. We show that the analysis can be phrased
purely in terms of local geometry at and near the wormhole throat,
and do not have to make any technical assumptions about asymptotic
flatness or other global properties. A key aspect of the analysis
is the demonstration that time-dependent wormholes have {\em two}
throats, one for each direction through the wormhole, and that the
two throats coalesce only for the case of a static wormhole.  }

\medskip

{\em Introduction:} The fact that traversable wormholes are
accompanied by unavoidable violations of the null energy condition
(NEC) is perhaps one of the most central features of wormhole
physics~\cite{Morris-Thorne,MTY,Visser-Book}. The original proof of
the necessity for NEC violations at or near the throat of a
traversable wormhole was limited to the static spherically symmetric
Morris-Thorne wormhole~\cite{Morris-Thorne}, though it was rapidly
realized that NEC violations typically occurred in at least some
explicit examples of static non-symmetric~\cite{Visser-Examples} and
spherically-symmetric time-dependent~\cite{Visser-Surgery}
wormholes. A considerably more general proof of the necessity of NEC
violations was provided by the {\em topological censorship theorem} of
Friedman, Schleich, and Witt~\cite{FSW} though this theorem requires
many technical assumptions concerning asymptotic flatness and
causality conditions that limit its applicability.

We have recently adopted a different strategy and sought to develop
new general theorems concerning energy condition violations at and
near the throat of traversable wormholes by focusing attention only on
the local behaviour of the geometry at and near the throat, and
discarding all assumptions about symmetry, asymptotic behaviour, and
causal properties. (This strategy was inspired by the fact that there
are many classes of objects that we would meaningfully wish to call
wormholes that possess either trivial topology~\cite{Visser-Book} or do
not necessarily possess asymptotically flat
regions~\cite{HPS}.) Such a strategy first requires
that we develop robust and general definitions of what it means to be
a wormhole throat.

Recently, we have succeeded in developing such definitions and
theorems, first for the static but completely non-symmetric
case~\cite{Hochberg-Visser,Visser-Hochberg}, and secondly (as reported
in this Letter) for the completely general time-dependent
non-symmetric wormhole. Our central results are these:

(1) A time-dependent traversable wormhole will have {\em two} throats, one
throat being associated with each direction of travel through the
wormhole. These two throats coalesce into one in the case of a static
traversable wormhole but it is important to keep them distinguished in
the time-dependent case.

(2) For each one of these throats the NEC is either violated or on the
verge of being violated on the throat itself. (This is easy to prove.)

(3) For each one of these throats there will be an open region
(topologically open in one of the null hypersurfaces passing through
the throat) whose closure intersects the throat such that the NEC is
violated throughout the open region. (Proving this requires a little
more analysis.)

(4) For each one of these throats there will be an open interval such
that the transverse averaged NEC (transverse averaged over the throat
after it is pushed out in the appropriate null direction) is violated
throughout the open interval. (Proving this requires a little more
analysis.)

The proofs are sketched below, and are based on an extension of an
idea due to Page~\cite{Page} whereby wormhole throats are viewed as
anti-trapped surfaces in spacetime. Further technical details may be
found in~\cite{Hochberg-Visser-Dynamic}, which also extends the
present analysis to spacetimes including non-zero torsion.

We note that there are a number of published papers which {\em claim}
to construct wormholes without NEC violations. These claims are most
often based on semantic confusion, though sometimes actual
computational errors have crept in, and we shall discuss the situation
more fully in our conclusions.


{\em Basic Definitions: Null geodesic congruences.} Consider a compact
two-dimensional spacelike hypersurface (denoted $\Sigma$) embedded in
$(3+1)$-dimensional spacetime. (More precisely, a compact two
dimensional orientable manifold that is embedded in spacetime in a
two-sided and spacelike manner.) At each point on the hypersurface
there are two null vectors orthogonal to the hypersurface, and these
two null vectors can be extended to two null geodesic congruences
(vector fields $l_\pm$) that are well defined on an open neighborhood
of the hypersurface~\cite{Hawking-Ellis,Wald}. Coordinates on the
hypersurface will be denoted $x$, while the null geodesic congruences
can be used to set up null coordinates $u_\pm$ on the orthogonal
hyperplanes. The $\pm$ labels are often denoted ``ingoing'' and
``outgoing'' though these labels are prejudicial and in the case of
non-trivial topology are actually meaningless, the key point is that
there are two possible null directions to travel along.

For each one of these null geodesic congruences one can define expansion,
shear, and vorticity, in a manner completely analogous to standard
textbook discussions~\cite{Hawking-Ellis,Wald}. In particular, we have
a pair of Raychaudhuri equations, one for each null geodesic
congruence, so that:
\begin{equation}\label{E-Raychaudhuri}
\frac{d \theta_{\pm}}{d u_{\pm}}= -\frac{1}{2}{\theta_{\pm}}^2 - 
{\sigma^{\pm}}^{ab}\sigma_{\pm ab}
+ {\omega^{\pm}}^{ab}\omega_{\pm ab} - R_c^d \; l^c_{\pm} l_{\pm d}.
\end{equation}
We now define a wormhole throat by demanding that at a throat the
cross-sectional area of a bundle of light rays passing orthogonally
through the putative throat should be at a strict local non-zero
minimum. We implement this notion by first defining the hypersurfaces
$\Sigma(u_\pm(x))$ to be the spacelike hypersurfaces formed by taking
$\Sigma\equiv\Sigma(0)$ and pushing it out an affine distance
$u_\pm(x)$ along the null congruence $l_\pm$. Define the area in terms
of the two-metric $\gamma$ induced on the hypersurface $\Sigma$ by
\begin{equation}
A[\Sigma(u_\pm(x))] = 
\int_{\Sigma(u_\pm(x))} \sqrt{\gamma(x,u_\pm(x),u_\mp=0)} \; d^2 x.
\end{equation}
If $\Sigma$ is to be regarded as a throat, we must demand first
that $A[\Sigma] > 0$ and second that there exists some open
neighborhood ${\cal U}$ surrounding the zero function $u_\pm(x)=0$
such that on this open set
\begin{equation}
A[\Sigma(u_\pm)] \geq A[\Sigma].
\end{equation}
Dropping the $\pm$ to simplify notation this implies first that
$\delta A/\delta u|_{(u=0)}=0$, and secondly that $\delta^2A/\delta
u^2|_{(u=0)} \geq 0$. But the null variational derivatives of the
cross-sectional area can be related to the extrinsic curvature of the
hypersurface $\Sigma$, where $\Sigma$ is to be considered as a submanifold
of the null hypersurface generated by the null congruence passing
through $\Sigma$. That is, $\Sigma$ is a throat with respect to the
null congruence $l_\pm$ provided first that the expansion of the null
congruence vanishes everywhere on the throat (the extrinsic curvature
is zero)
\begin{equation}
\theta_\pm =0,
\end{equation}
and secondly that everywhere on the throat the expansion of the null
congruence satisfies the (simple) ``flare out'' condition
\begin{equation}
{d\theta_\pm\over du_\pm} \geq 0.
\end{equation}
As it stands, this preliminary and simple definition of flare-out
just marginally fails to distinguish a throat from a trapped surface,
and so the equations (\ref{Ricci}) and (\ref{stress}) apply to
throats as well to apparent horizons (and apply in fact to arbitrary
cross sections of the apparent horizon of a dynamic black hole).
We refine and strengthen our flare-out definition below so as to
exclude apparent horizons.  This preliminary definition is already
enough to prove the first two of our key results.


{\em Theorems:}
First we note that a hypersurface that is extremal with respect to one
of the null congruences (say $l_+$) will in general not be extremal with
respect to the other. This is particularly clear in the case of a
spherically symmetric but time-dependent wormhole where the two radial
null geodesic congruences are clearly of paramount importance. Looking
for the zero of $\theta_+$ defines the wormhole throat for light
travelling in the $l_+$ direction, looking for the zero of $\theta_-$
defines the wormhole throat for light moving in the $l_-$
direction. In the special case where the wormhole is static (not
necessarily spherically symmetric) the two null vectors can be
decomposed as $l_\pm = (V \pm n)/2$, where $V$ is a unit vector
parallel to the time-translation Killing vector and $n$ is a unit
spacelike vector normal to $\Sigma$ and orthogonal to $V$. With this
notation the expansion can be computed in terms of the Lie derivative
of the induced two-tensor on $\Sigma$ and is seen to be
\begin{eqnarray}\label{Lie}
\theta_\pm 
&=&
 {1\over2} \Tr\left[{\cal L}_{l_\pm} \gamma\right]
\nonumber\\
&=&
 {1\over4} \left( \Tr\left[{\cal L}_V \gamma\right] 
                  \pm \Tr\left[{\cal L}_n \gamma\right] \right)
\nonumber\\
&=&
 \pm {1\over4}  \Tr\left[{\cal L}_n \gamma\right].
\end{eqnarray}
This is completely in agreement with our previous analysis for the
static case~\cite{Hochberg-Visser,Visser-Hochberg} and, since the
vanishing of $\theta_+$ now automatically implies the vanishing of
$\theta_-$ and vice versa, shows that in the static case the two
throats coalesce into one.

Note that the two throats defined above share many of the
properties more typically attributed to apparent
horizons~\cite{Hawking-Ellis,Wald}. The throats can move around as
a function of time in a rather complicated fashion and the
three-surface swept out by the throats as a function of time need
not necessarily be smooth, nor need this three-surface necessarily
have a spacelike normal, though typically it will.
Moreover, the relative motion of the two throats can be such
as to render the wormhole either one-way traversable or two-way
traversable (meaning from one throat to the other), 
depending on the degree of causal connectedness
of the two three-surfaces $W(\Sigma_{\pm})$ swept out by the two throats. 
That is, one may have $I^+(\Sigma_{+}) \bigcap W(\Sigma_{-}) \neq \phi$,
and/or $I^+(\Sigma_{-}) \bigcap W(\Sigma_{+}) \neq \phi$ or perhaps neither
condition holding, where $I^+$ denotes the chronological future.

Second, by applying the Raychaudhuri equation, using the fact that the
twist $\omega_\pm$ is automatically zero for any orthogonal geodesic
congruence, that $\sigma^2 \geq 0$, and noting that by definition
$\theta_\pm=0$ and $d\theta_\pm/du_\pm \geq 0$ at the throat, we
immediately derive
\begin{equation}\label{Ricci}
R_{ab} \; l_\pm{}^a l_\pm{}^b \leq 0,
\end{equation}
where we must use the {\em same} null vector as is used in defining the
throat. Applying the Einstein equations now yields
\begin{equation}\label{stress}
T_{ab} \; l_\pm{}^a l_\pm{}^b \leq 0,
\end{equation}
so that the NEC is either violated or on the verge of being violated
at the throat ($T_{ab} \; l_\pm{}^a l_\pm{}^b$ could be zero). Of
course we want to derive a stronger result that replaces the weak
inequality above by a strict inequality. Proving this is a little
tedious and requires just a bit of mathematical analysis.

We supplement the preliminary definition of a throat given above with the
condition that there be at least {\em some} (infinitesimal) variations
for which the area is strictly increasing. More precisely suppose that
there exists at least one  $u_\pm(x)=\epsilon f(x)$ with $\epsilon\in
(-\delta,0)\cup (0,+\delta)$ for which the area function satisfies the
strict inequality
\begin{equation}
A[\Sigma(u_\pm=\epsilon f)] > A[\Sigma].
\end{equation}
(As discussed more fully
in~\cite{Hochberg-Visser,Visser-Hochberg,Hochberg-Visser-Dynamic},
this constraint disposes of certain degenerate cases and guarantees
that $\Sigma$ is a true local minimum of the area: there is at least
one direction in which the area actually increases.) Application of
the fundamental theorem of calculus now implies that there is an open
interval in $\epsilon$ (in general $(-\delta_2,0)\cup (0,+\delta_2)$,
different from that above) for which
\begin{equation}
{d^2A[\Sigma(u_\pm=\epsilon f)]\over d\epsilon^2} > 0,
\end{equation}
with this now being a {\em strict} inequality.  Apparent horizons
and bifurcation two-spheres fail to satisfy these strict inequalities
and so are {\it not} examples of throats.  In terms of the expansion
this implies the existence of an open interval in $\epsilon$ (in
general $(-\delta_3,0)\cup (0,+\delta_3)$, different from that
above) for which
\begin{equation}
\label{E-integrated}
\int_{\Sigma(u_\pm=\epsilon f)} 
d^2 x \, \sqrt{\gamma} \, f^2(x) 
{d\theta_\pm\over du_\pm} > 0,
\end{equation}
This constraint on the expansion further implies the existence of
another open neighborhood, this time in the appropriate null
hypersurface passing through $\Sigma$ such that
\begin{equation}
\left.{d\theta_\pm\over du_\pm}\right|_{(x,u_\pm,u_\mp=0)} > 0.
\end{equation}
This last open set may include the hypersurface $\Sigma(0)$ itself
but does not necessarily have to do so, its closure must however
certainly intersect $\Sigma(0)$. These messy technical details are
required to justify replacing the weak inequality by a strict
inequality and occur in some form or another in all technical
discussions of NEC violations in traversable wormholes. It is
because of these technical complications that the phrase ``at or
near the throat'' is ubiquitous.  (See in particular the discussions
in~\cite{Hochberg-Visser,Visser-Hochberg,Hochberg-Visser-Dynamic}.)

Of course, with these incantations out of the way, the proof of NEC
violations follows along the lines indicated previously and we find
that there is an open neighborhood near the throat over which
\begin{equation}
T_{ab} \; l_\pm{}^a l_\pm{}^b < 0.
\end{equation}
This completes the proof of the third key result enunciated above,
while the fourth result follows from appropriate modifications of
the integrated constraint Eq. (\ref{E-integrated}).
See~\cite{Hochberg-Visser-Dynamic,sgn} for details.


{\em Discussion:} We have thus shown that the violations of the
NEC commonly ascribed to traversable wormholes are completely
generic --- NEC violations will occur at or near the throat of any
spacetime configuration that deserves to be called a (traversable)
wormhole, and this result is completely general and independent of
issues of symmetry, asymptotic flatness, or time dependence.  The
present discussion is completely in agreement with the original
Morris-Thorne analysis~\cite{Morris-Thorne}, the topological
censorship theorem~\cite{FSW}, and our own earlier analyses of the
generic static wormhole~\cite{Hochberg-Visser,Visser-Hochberg}.

On the other hand, the striking nature of these NEC violations has led
to a minor industry of papers claiming to construct wormholes without
violating the energy conditions. We can classify these attempts into
three classes: (1) playing semantic games, (2) genuine ambiguities,
and (3) simple error.

One way of playing semantic games is to arbitrarily divide the
total stress energy into an ``exotic component'' plus a ``normal
component''. As seen above, the total stress energy must always
violate the NEC, but it is sometimes possible to force all the NEC
violations into the ``exotic component'' of the stress-energy and
keep the ``normal component'' well-behaved.  This strategy of
semantic confusion underlies all the claims that Brans-Dicke
wormholes do not violate the NEC, since for suitable choice of the
$\omega$ parameter the Brans-Dicke scalar can be coerced into
providing the NEC violations.  (Of course, the fact that for certain
values of the $\omega$ parameter Brans-Dicke gravity supports
traversable wormhole solutions is both interesting and non-trivial,
it is only the claimed lack of NEC violations that we find
problematical.) Some papers adopting this strategy
are~\cite{Brans-Dicke-Papers}, and we have provided a fuller
treatment of the situation in~\cite{Visser-Hochberg}. The same sort
of comments should be borne in mind for wormholes based on other
variants of Einstein gravity, whether they be higher-derivative
gravity~\cite{Kar3}, dilaton gravity, scalar-tensor gravity, or
whatever. (We mention that adding torsion actually makes the problem
worse not better, see~\cite{Hochberg-Visser-Dynamic}.)

Another technique commonly used is to rely on time dependence to
temporarily and locally suspend the violations of the NEC.
Time-dependent wormholes in an inflationary context were first
studied by Roman~\cite{Roman} (see also \cite{Hochberg-Kephart}),
who correctly recognized the existence of NEC violations. Subsequent
papers have sought to utilize time-dependence to evade the NEC
violations~\cite{Kar1,Kar2,Wang-Letelier,Time-dependent}, and some
commentary regarding these attempts can be found
in~\cite{Visser-Hochberg,Hochberg-Visser-Dynamic}.  The key here
is to note that the suspension of the NEC is essentially an illusion
in that if one ever succeeds in passing through the wormhole and
reaching the other universe then one must have passed a throat as
defined above and there must be NEC violations at or near this
throat. The NEC violations can be moved around to some extent, but
if the NEC violations are suspended, then the ingoing null congruence
is still contracting.  A subtlety that has helped cause confusion
in the past is that (as we have seen in this Letter) time-dependent
wormholes typically have two throats. If the wormhole is symmetric
under interchange of the two universes it connects, then these two
throats will also be symmetrically placed but they will {\em not}
be at the center of symmetry of the wormhole. It is only for static
wormholes that the two throats coalesce.  (This fact makes it easy
to get confused about what it means to ``pass through'' a time-dependent
wormhole --- it is not enough to merely pass the central symmetry
point. In passing through an expanding wormhole, the throat [extremal
area] is encountered before reaching the center, while for a
contracting wormhole the throat is encountered after crossing the
center.)

Another source of confusion associated with past analyses of
time-dependent wormholes arises from the use of spatial embedding
techniques for defining and locating throats.  These techniques
require selecting and lifting out a particular time-slice from the
given spacetime believed to contain a wormhole and then embedding this
instantaneous three-geometry in a flat Euclidean ${\cal R}^n$.
For a static wormhole, any constant time-slice will suffice
and flare-out in the spatial embedding direction orthogonal to the
throat is sufficient to imply flare-out in spacetime.  But for a
dynamic wormhole, where the two-dimensional throats sweep out
complicated extended three-dimensional objects in spacetime,
flare-out in the embedding direction (typically spacelike) does
not imply flare-out in the {\it null} directions orthogonal to the
two-dimensional throat(s).  (Computation has also convinced us that
direct search for some extremal three-surface with spacelike normal
is useless for deriving NEC violation theorems---the technical
deficiency of such an approach is that without a natural definition
of the ``time'' direction it is impossible to define what one means
by null vectors orthogonal to a timelike three-surface.) Simple
spatial embedding constructions also completely miss detecting the
presence of the {\em two} throats.

Finally, we mention that a distressing number of papers addressing
this topic are marred by actual calculational error, and encourage
interested readers to proceed with caution.

In summary, we feel that the original observation by Morris and Thorne
that traversable wormholes are accompanied by NEC violations has now
been successfully and completely generalized to arbitrary traversable
wormholes. The subtleties involved in the generalization to the
time-dependent case lie in the fact that there are in general two
wormhole throats for time-dependent traversable wormholes, and that
some careful technical steps must be taken in the analysis in order to
get the desired strong inequality (a weak inequality is much easier to
derive). For generic static wormholes, the Gauss-Codazzi decomposition
of the curvature tensor enables us to make considerably more detailed
statements as to the curvature of spacetime near the throat.


{\em Acknowledgements:}
This work was supported by Spain's National Aerospace Technology
Institute (INTA) (D.H.), and by the US Department of Energy (M.V.).
M.V. acknowledges interesting and constructive comments made
by Sean Hayward.



\begin{references}
\bibitem{sgn} The averaged flare-out condition places a constraint
on the putative throat by asking that the extremal surface be
outward-flaring over at least half its area and must be carefully
formulated to remain invariant under arbitrary affine reparametrizations
of the null geodesic congruence. See \cite{Hochberg-Visser-Dynamic}
for more details.
\end{references}
\end{document}